\documentclass[conference]{IEEEtran}
\IEEEoverridecommandlockouts
\usepackage{cite}
\usepackage{amsmath,amssymb,amsfonts}
\usepackage{algorithmic}
\usepackage{graphicx}
\usepackage{textcomp}
\usepackage{xcolor}
\usepackage{subcaption}
\def\BibTeX{{\rm B\kern-.05em{\sc i\kern-.025em b}\kern-.08em
    T\kern-.1667em\lower.7ex\hbox{E}\kern-.125emX}}
\begin{document}

\title{Low-Energy and CPA-Resistant Adiabatic CMOS/MTJ Logic for IoT Devices}

\author{\IEEEauthorblockN{Zachary Kahleifeh and Himanshu Thapliyal }
\IEEEauthorblockA{VLSI Emerging Design And Nano Things Security Lab (VEDANTS-Lab) \\ Department of Electrical and Computer Engineering, University of Kentucky, Lexington, KY, USA\\
Email:\space hthapliyal@uky.edu}}

\maketitle

\begin{abstract}
The tremendous growth in the number of Internet of Things (IoT) devices has increased focus on the energy efficiency and security of an IoT device. In this paper, we will present a design level, non-volatile adiabatic architecture for low-energy and Correlation Power Analysis (CPA) resistant  IoT devices. IoT devices constructed with CMOS integrated circuits suffer from high dynamic energy and leakage power. To solve this, we look at both adiabatic logic and STT-MTJs (Spin Transfer Torque Magnetic Tunnel Junctions) to reduce both dynamic energy and leakage power. Furthermore, CMOS integrated circuits suffer from side-channel leakage making them insecure against power analysis attacks. We again look to adiabatic logic to design secure circuits with uniform power consumption, thus, defending against power analysis attacks. We have developed a hybrid adiabatic-MTJ architecture using two-phase adiabatic logic. We show that hybrid adiabatic-MTJ circuits are both low energy and secure when compared with CMOS circuits. As a case study, we have constructed one round of PRESENT and have shown energy savings of 64.29\% at a frequency of 25 MHz. Furthermore, we have performed a correlation power analysis attack on our proposed design and determined that the key was kept hidden. 
\end{abstract}

\begin{IEEEkeywords}
Adiabatic Logic, Magnetic Tunnel Junction, Correlation Power Analysis, Side-channel Attacks, Internet of Things (IoT). 
\end{IEEEkeywords}

\section{Introduction}
The age of portable devices has resulted in a sharp upward trend in Internet of Things (IoT) devices. IoT devices are typically battery-operated, and thus the need for energy-efficient processors is high. Furthermore, many IoT devices store and transmit sensitive data and thus the security of IoT devices should not be neglected \cite{alioto2017internet}. In this paper, we look to create a secure device against power analysis attacks without suffering from energy efficiency degradation. To remain secure against power analysis attacks and consume lower energy we look to both non-volatile memory in the form of Spin Transfer Torque Magnetic Tunnel Junction (STT-MTJ) \cite{huai2008spin} and a low energy design technique known as adiabatic logic. 

STT-MTJ has numerous advantages over common memory technologies such as extremely low standby power, non-volatility, easy compatibility with CMOS, and high integration density \cite{deng2013low, kang2016low, kang2015spintronics}. MTJs can be combined with standard CMOS devices to create low-energy circuits \cite{zhao2013synchronous}. 

\begin{figure*}
	\centering
	\includegraphics[width=0.62\linewidth]{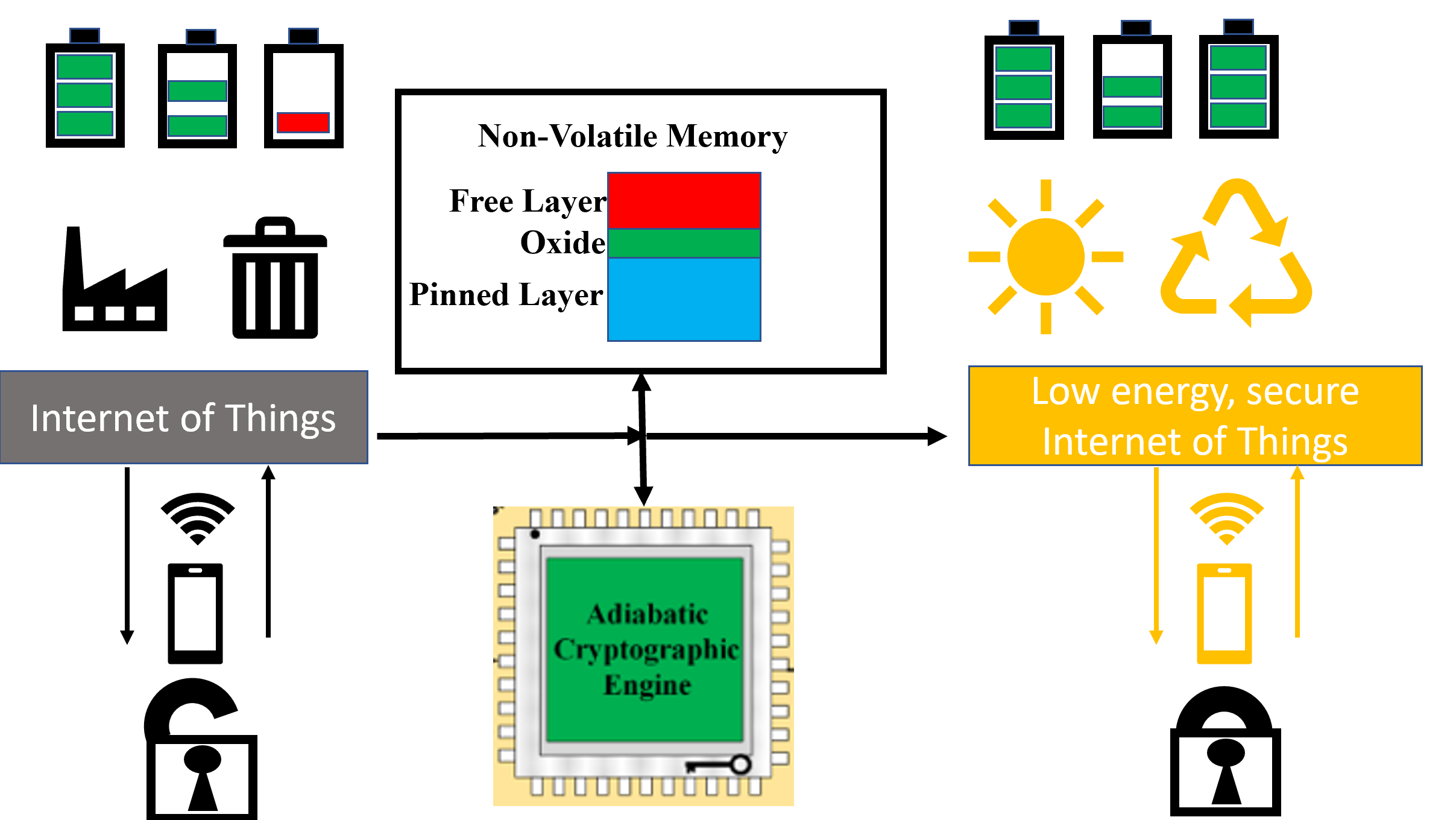}
	\caption{Hybrid adiabatic-MTJ circuits can introduce a golden age to IoT devices.}
	\label{fig:ill}
\end{figure*}

While MTJ based circuits reduce standby power, adiabatic-based circuits can reduce overall energy consumption. Adiabatic logic is an emerging design technique to design low energy and secure circuits. Adiabatic logic recycles energy from the load capacitor back into the clock generator to reduce energy consumption \cite{teichmann2011adiabatic}.

When reducing the energy consumption of a device, security should not be neglected. The threat vector of IoT devices continues to grow thus defenses against these attacks should be developed. One such security threat that IoT devices can experience is a class of hardware attacks known as side-channel attacks. Side-channel attacks look to retrieve hidden information through a device's side-channel such as power consumption\cite{kocher1999differential}, circuit timing \cite{dhem1998practical}, etc. Side-channel attacks are a dangerous threat to device functionality and vital device information such as encryption keys. One particular side-channel attack we will focus on is the Correlation Power Analysis Attack (CPA) \cite{wu2012measurement}. This attack looks to correlate power with bits to retrieve hidden information.

\begin{figure}
	\centering
	\includegraphics[width=0.7\linewidth]{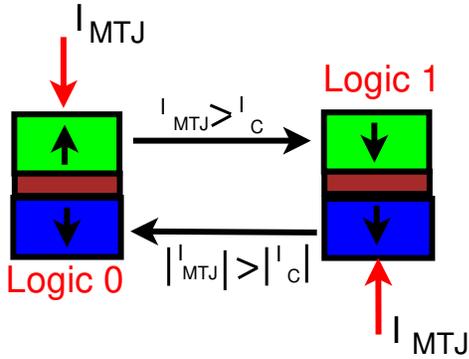}
	\caption{Structure of Magnetic Tunnel Junction with Spin Transfer Torque (STT) switching.}
	\label{fig:MTJnanopillar}
\end{figure}

In this paper, we look to combine the emerging technology of STT-MTJs with the emerging design technique of adiabatic logic to design ultra-low energy circuits while also remaining secure against power analysis attacks. To demonstrate energy savings, we have constructed one round of the PRESENT encryption algorithm using our proposed hybrid adiabatic-MTJ architecture \cite{bogdanov2007present}. Our simulations show that when compared with CMOS our designs save 64.29\% at 25 MHz. To demonstrate secure operations, we also performed a CPA attack on the PRESENT Substitution Box (S-Box). When performing the attack on the CMOS implementation we were able to retrieve the secret encryption key. However, when performing the attack on our hybrid adiabatic-MTJ design we were not able to steal the key thus demonstrating its resilience against CPA attacks.

\section{Background} 

\subsection{Magnetic Tunnel Junction}
Magnetic Tunnel Junction (MTJ) consists of two ferromagnetic (FM) layers and an oxide layer that serves as a barrier between the two ferromagnetic layers \cite{moodera1995large}. The magnetization of one of the FM layers is fixed in most circuit applications of MTJs, while the other FM layer is free to take either a parallel or antiparallel magnetization \cite{zand2016scalable}. This can be seen in Figure \ref{fig:MTJnanopillar} as the bottom layer of the MTJ is fixed and the top layer is free to take a direction. If the MTJ shows a parallel magnetization ($R_{P}$) then it will have lower resistance than when it has an antiparallel magnetization ($R_{AP}$). \cite{behin2014computing}. The MTJ structure and two configurations are shown in Figure \ref{fig:MTJnanopillar}. The difference in resistance between the two states of the MTJ devices is given by the tunnel magnetoresistance ratio $TMR=(R_{AP}-R_{P})/R_{P}$. MTJ devices with higher TMR ratios have been shown to have higher reliability \cite{kent2010perpendicular}.

\subsection{CMOS-MTJ Hybrid Circuits} 
\begin{figure}
	\centering
	\includegraphics[width=\linewidth]{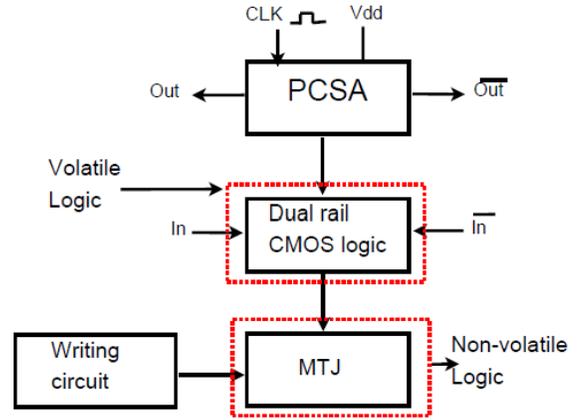}
	\caption{General structure of Hybrid CMOS-MTJ circuits.}
	\label{fig:cmos_mtj}
\end{figure}

Figure \ref{fig:cmos_mtj} shows the general structure of an existing version of a  Logic-In-Memory (LIM) based CMOS-MTJ circuit. The LIM architecture consists of a Pre-Charged Sense Amplifier (PCSA) circuit consisting of MP1, MP2, MN1, and MN2. A dual-rail NMOS only logic tree (T1-T4) evaluates the inputs and the non-volatile MTJs store data. The write circuit is used to switch the MTJs when the respective input is switched.


The operation of the PCSA can be explained through the existing PCSA based CMOS/MTJ XOR gate (Figure \ref{fig:xor})\cite{deng2013low} \cite{gang2011high}. The PCSA, which uses a CLK signal, operates in two phases. The outputs are pre-charged to "1" when CLK is set to "0" and the output voltages begin to discharge to ground when CLK is set to "1". The discharge speed will be different for each branch due to the difference in resistance of the different MTJ configurations (parallel and antiparallel). For example, if MTJ1 is configured in parallel mode and MTJ2 is configured in antiparallel mode, then $R_{MTJ2} > R_{MTJ1}$. Due to the difference in resistances between $R_{MTJ1}$ and $R_{MTJ2}$, the discharge current through MTJ1 will be greater than MTJ2. When XOR reaches the threshold voltage of MP1, XNOR will be charged to “1” and XOR will be discharged to “0”.

\begin{figure}
	\centering
	\includegraphics[width=\linewidth]{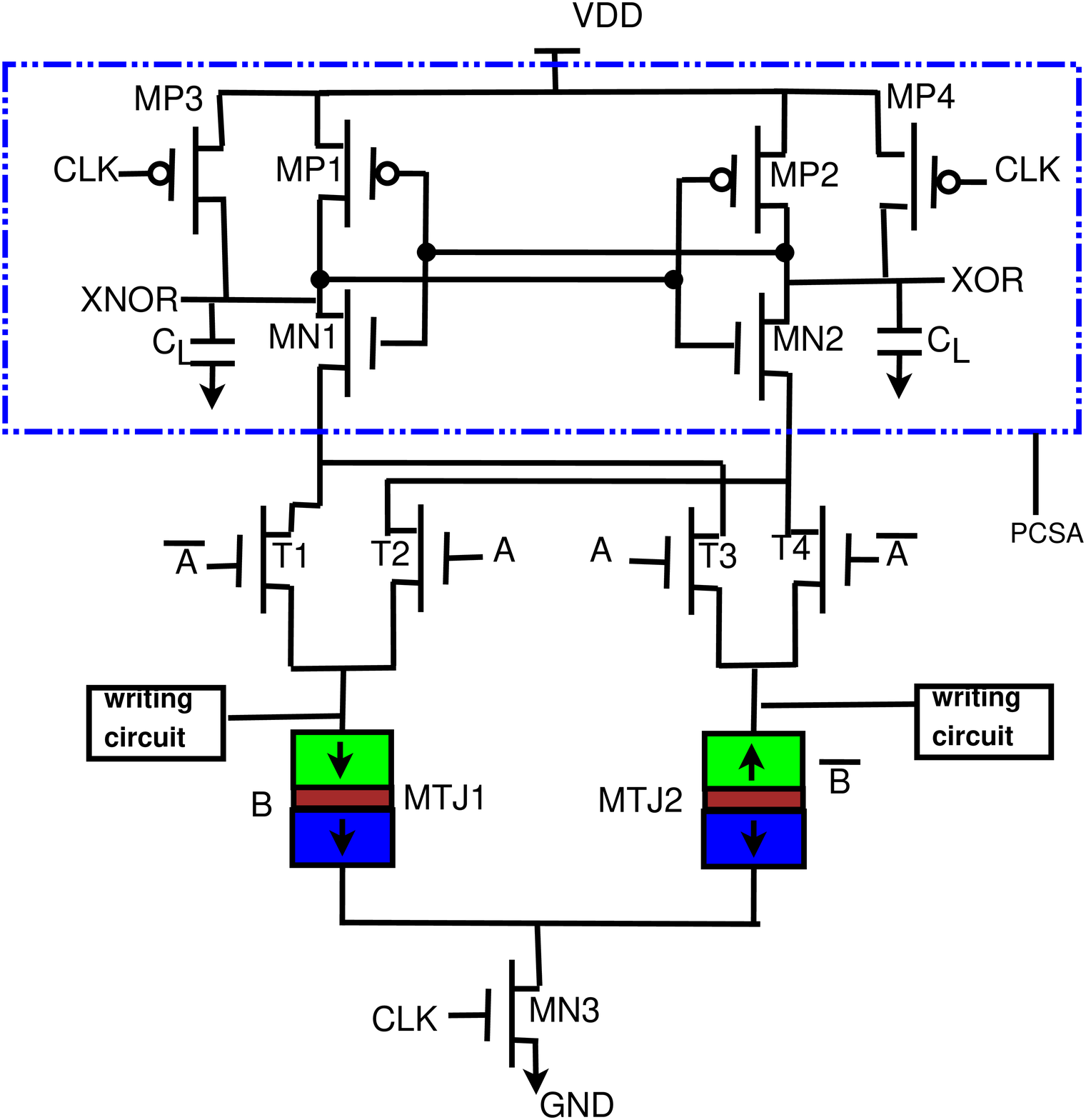}
	\caption{Hybrid CMOS-MTJ XOR circuit \cite{deng2013low}\cite{gang2011high}.}
	\label{fig:xor}
\end{figure}

\subsection{Adiabatic Logic} 

\begin{figure}
\centering

  \includegraphics[width=\linewidth]{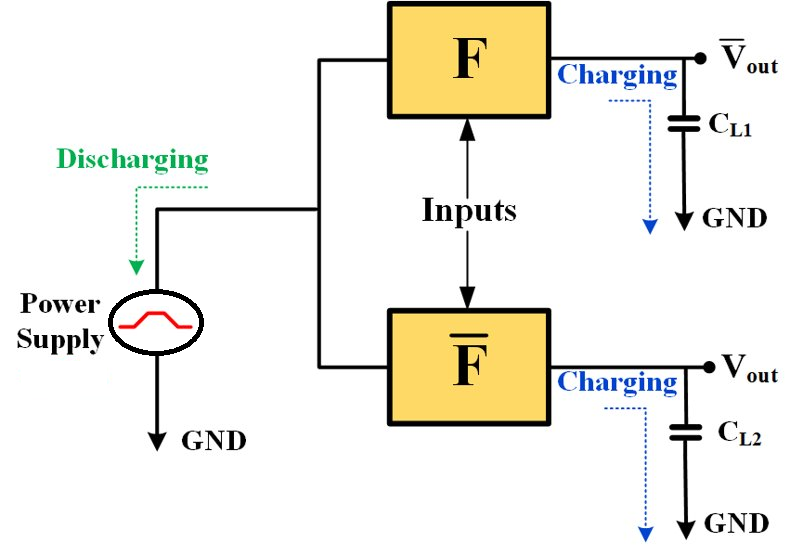}
\caption{Adiabatic charging and recovery principle.}
\label{fig:charging}       
\end{figure}

Adiabatic logic is a circuit design technique for designing ultra-low-energy circuits \cite{athas1994low}. Adiabatic logic
reduces the energy of a circuit by recovering the energy stored in the load capacitor at the end of each clock cycle. The recovered energy is stored either through magnetic energy in clock inductors or through an electric charge in the clock capacitance. The recovered energy is then reused in the next cycle to reduce energy consumption.
The energy dissipated in an adiabatic circuit is given by: 
\emph{}\begin{equation}
\label{eqn5}
E_{diss} = \frac{RC}{T}CV_{dd}^{2}
\end{equation}
Where $T$ is the charging period of the capacitor, $C$ is the output load
capacitor, $V_{dd}$ is the full swing of the power clock. If the charging time $T>2RC$, then the energy dissipated by an adiabatic circuit is less than a conventional CMOS circuit. Figure \ref{fig:charging} illustrates the principle of energy recovery within an adiabatic system. 

\section{Proposed Secure Hybrid Adiabatic-MTJ  Circuit}
\begin{figure}
\centering

  \includegraphics[width=\linewidth]{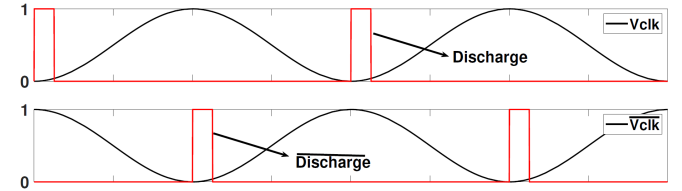}
\caption{CPA-resistant two-phase adiabatic logic clocking scheme.}
\label{fig:clk}       
\end{figure}

In this section, we will review the structure of our proposed hybrid adiabatic-MTJ circuit. The proposed XOR/XNOR gate circuit can be seen in Figure \ref{fig:xor}. We can see that the structure consists of a 2 PMOS and 2 NMOS (2P2N) Pre-Charged Sense Amplifier (PCSA). There is also a dual-rail evaluation network that consists of only NMOS transistors connected to two MTJs with opposite configurations. Finally, two NMOS transistors are used to discharge the outputs before the next clock cycle begins. Our proposed hybrid adiabatic-MTJ uses a two-phase clocking scheme consisting of two sinusoidal clocks 90$^{\circ}$ out of phase as well as two discharge signals in phase with the respective clocks. The clocking waveform for two-phase adiabatic logic can be seen in Figure \ref{fig:clk}.

\begin{figure}
\centering

  \includegraphics[width=\linewidth]{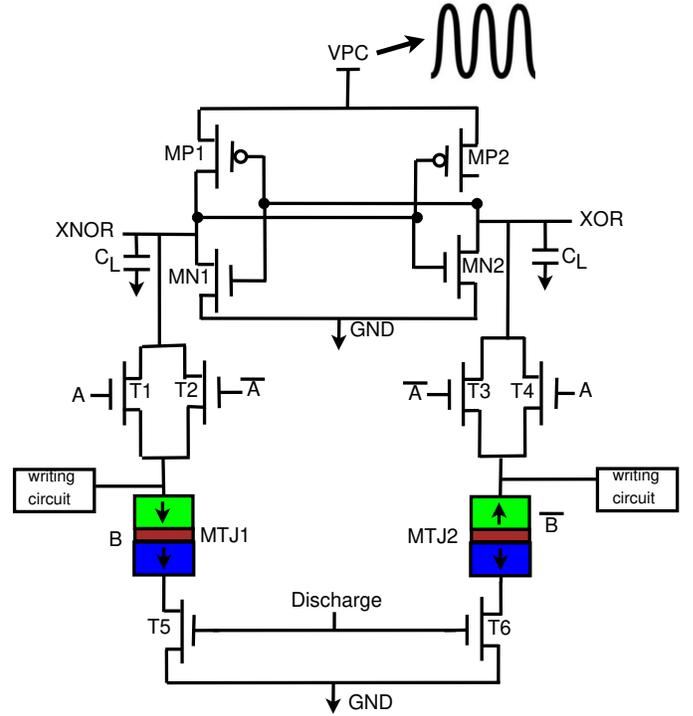}
\caption{Proposed low energy and secure adiabatic-MTJ XOR/XNOR gate.}
\label{fig:xor}       
\end{figure}

\subsection{Proposed Hybrid Adiabatic-MTJ PRESENT Implementation} 
PRESENT \cite{bogdanov2007present} is an ultra-lightweight block cipher. PRESENT has low area when compared with other block ciphers which makes it a strong choice for implementation in area constrained IoT devices that look to be resilient against CPA attacks. In this paper, we intend to use the 80-bit version of PRESENT. 

\begin{figure}
\centering	

  \includegraphics[width=\linewidth]{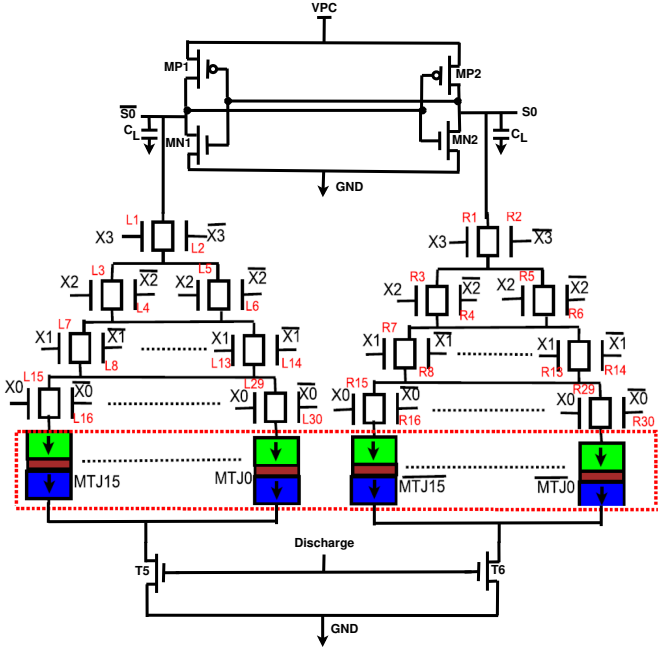}
\caption{Proposed hybrid adiabatic-MTJ S-box LUT.}
\label{fig:sbox}       
\end{figure}

One of the components of PRESENT is the substitution box (S-box) which performs a non-linear substitution. When constructed with CMOS, the S-box consumes high energy and is prone to Correlation Power Analysis Attacks (CPA) thus we look to construct the S-box using our hybrid adiabatic-MTJ circuit. When constructing the S-box circuit using our proposed design we intend to limit the switching of MTJs to reduce energy consumption. To do this, we construct the S-box using a Look-Up-Table (LUT) based method in which the MTJs are written only once so that the output of the S-box is stored within the MTJs. Figure \ref{fig:sbox} illustrates our proposed S-box.

\begin{figure}
\centering	

  \includegraphics[width=\linewidth]{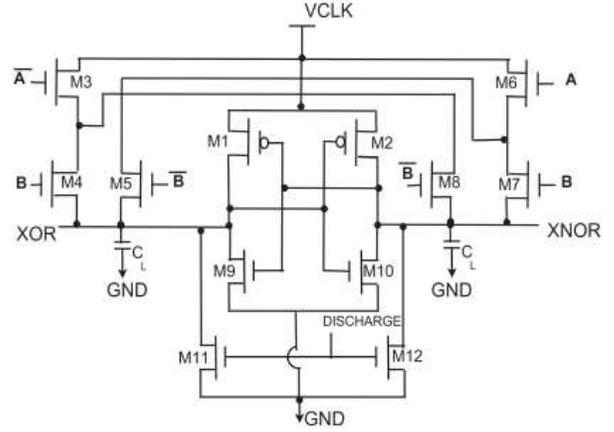}
\caption{2-EE-SPFAL XOR Gate used to implement PRESENT.}
\label{fig:sp_xor}       
\end{figure}

\begin{figure*}
\centering	

  \includegraphics[width=\textwidth]{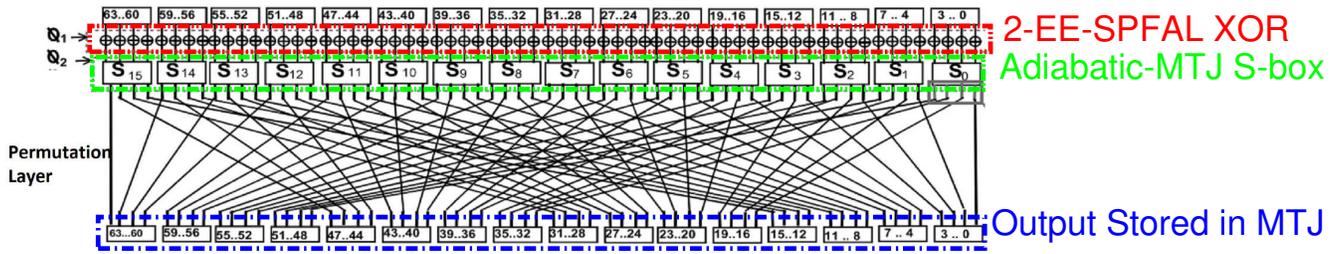}
\caption{Complete structure of 1-Round of PRESENT implemented with 2-EE-SPFAL \cite{kahleifeh20202} and hybrid Adiabati-MTJ.}
\label{fig:present}       
\end{figure*}

Another component of PRESENT is the XOR gate. As mentioned previously, MTJ circuits consume high power when there is frequent switching thus the XOR circuit cannot be designed with the proposed adiabatic-MTJ circuit. Instead, we have designed our XOR gate using 2-EE-SPFAL \cite{kahleifeh20202}. 2-EE-SPFAL has been shown to be CPA-resistant and low energy which allows the implementation of PRESENT to also be secure and low-energy. The complete implementation of PRESENT can be seen in Figure \ref{fig:present}. The XOR gate can be seen in Figure \ref{fig:sp_xor}.

\section{Simulation Results}
The simulation results of the proposed hybrid adiabatic-MTJ based circuits are presented in this section. Simulations are performed using Cadence
Spectre simulator with 45nm standard CMOS technology with perpendicular anisotropy CoFeB/MgO MTJ model \cite{wang2016compact}. Because we do not switch our MTJs, we model the device using a basic resistor. The resistance values are calculated based on our MTJ parameters which are listed in Table \ref{tab:param}.

\begin{table}
\caption{NED and NSD values for hybrid adiabatic-MTJ S-box. }
\label{tab:param}       
\begin{tabular}{lll}
\hline\noalign{\smallskip}
Parameter & Description & Value  \\
\noalign{\smallskip}\hline\noalign{\smallskip}
$t_{sl}$ & Thickness of free layer & 1.3nm\\
a & Length of surface long axis & 40nm\\
b & Width of surface short axis & 40nm\\
$t_{ox}$	& Thickness of the Oxide barrier & 0.85nm\\
TMR & 0.Tunnel Magneto Resistance ratio & 150\%\\
RA & Resistance Area Product & $5\Omega \mu^2$\\
Area & MTJ layout surface & 40nm x 40nm x $\pi$/4\\
$R_p$ & Parallel resistance & 6.21 k$\Omega$\\
$R_{ap}$ &Antiparallel resistance  & 18.64 k$\Omega$\\ 
\noalign{\smallskip}\hline
\end{tabular}
\end{table}

\subsection{Normalized Energy Deviation and Normalized Standard Deviation}

The two criteria we will use to evaluate the security of our proposed design are Normalized Energy Deviation and Normalized Standard Deviation. The criteria Normalized Energy Deviation (NED) is defined as ($E_{max}$ - $E_{min}$)/$E_{max}$. NED is used to determine the percent difference between the minimum and maximum energy consumption. A second parameter, Normalized Standard Deviation (NSD), is defined as $\frac{\sigma_{e}}{\overline{E}}$ where $\sigma_{e}$ is the standard deviation of the energy dissipated by the circuit per input transition and $\overline{E}$ is the average energy dissipation. Both NED and NSD are important criteria when determining circuit resilience to CPA attacks. The lower the NED and NSD value the more uniform the power consumption and therefore the more secure a circuit is.

In this paper, we have calculated the NED and NSD values for our proposed S-box. Table \ref{tab:sbox} shows the NED and NSD values for our proposed design as well as a standard CMOS-based S-box as a base value to compare. From Table \ref{tab:sbox} we can see that our proposed design has lower average energy consumption than the CMOS-based S-box. Furthermore, our proposed S-box has lower NED and NSD values pointing towards its ability to defend against power analysis attacks. 
\begin{table}
\centering
\caption{NED and NSD values for hybrid adiabatic-MTJ S-box. }
\label{tab:sbox}       
\addtolength{\tabcolsep}{-0pt}
\begin{tabular}{lll}
\hline\noalign{\smallskip}
Parameter & Proposed S-box & CMOS \\
\noalign{\smallskip}\hline\noalign{\smallskip}
$E_{min} (fJ)$ & 33.7 & 7.1\\
$E_{max} (fJ)$ & 34.0 & 102.0\\
$E_{avg} (fJ)$ & 33.9 & 54.8\\
NED(\%)	& 0.80 & 93.0\\
NSD(\%) & 0.18 & 42.0\\
\noalign{\smallskip}\hline
\end{tabular}
\end{table}

\subsection{Hybrid Adiabatic-MTJ Case Study: 1-Round of PRESENT}
\begin{figure}
\centering	

  \includegraphics[width=\linewidth]{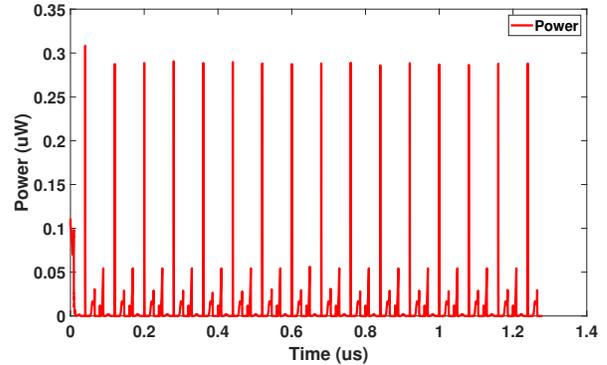}
\caption{Proposed hybrid adiabatic-MTJ S-box LUT uniform power consumption.}
\label{fig:current}       
\end{figure}

\begin{figure}
\centering	

  \includegraphics[width=\linewidth]{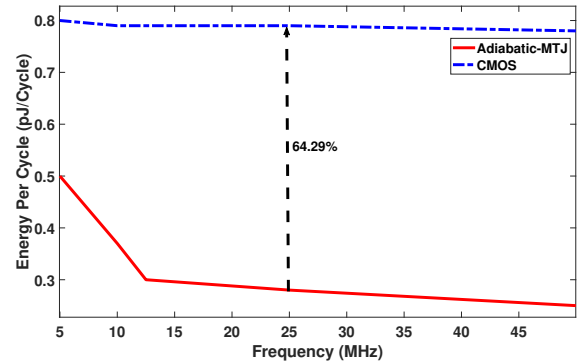}
\caption{Energy per cycle of hybrid adiabatic-MTJ and CMOS implementations of PRESENT.}
\label{fig:energy}       
\end{figure}

A PRESENT S-box implemented with a hybrid adiabatic-MTJ circuit consumes uniform power and is therefore secure against Correlation Power Analysis (CPA) Attacks. Uniform power consumption from 1 round of PRESENT can be seen in Figure \ref{fig:current}. The uniform power consumption is an indicator that the circuit is secure against a CPA attack as we will see when one is performed. MTJs are non-volatile memories that store data within the MTJs therefore, as a fair comparison we have added 64 Flip-Flops to the CMOS implementation to synchronize the inputs and store the output. Figure \ref{fig:energy} and Table \ref{table:en_per_cyc}  shows the energy per cycle of the proposed hybrid adiabatic-MTJ circuit and CMOS implementations of 1 round of PRESENT. From Figure \ref{fig:energy} and Table \ref{table:en_per_cyc} we can see that our proposed design consumes 0.50 pJ/cycle at 5 MHz while the CMOS implementation consumes 0.80 pJ/cycle resulting in a 36.6\% reduction in energy. At 50 MHz, our proposed design consumes 0.25 pJ/cycle while the CMOS implementation consumes 0.78 pJ/cycle resulting in a substantial energy reduction of 67.2\%. 

\begin{table}[!h]
\centering 
\addtolength{\tabcolsep}{-5pt}
\caption{Energy per cycle of Present-80 implemented with CMOS and hybrid Adiabatic-MTJ}
\begin{tabular}{| c | c | c | c | c | c | } 
\hline
Energy Per Cycle (pJ/Cycle) & 5 MHz & 10 MHz & 12.5 MHz & 25 MHz & 50 MHz \\
\hline
CMOS & 0.80 & 0.79 & 0.79 & 0.78 & 0.78  \\
\hline
Adiabatic-MTJ & 0.50 & 0.37 & 0.30 & 0.28 & 0.25  \\
\hline
Energy Reduction (\%) & 37.6 & 53.5 & 61.8 & 64.2 & 67.2\\
\hline
\end{tabular}
\label{table:en_per_cyc}
\end{table}

\subsection{CPA Attack on PRESENT-80}
\begin{figure}
\centering
\begin{subfigure}{\linewidth}
\centering
 \includegraphics[width=\linewidth]{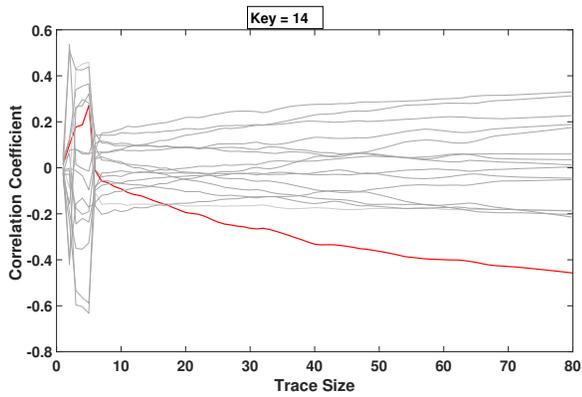}
 \caption {Successful CPA attack on CMOS based implementation of PRESENT S-box.}
\label{cmos_cpa}
\end{subfigure}

\begin{subfigure}{\linewidth}
\centering
  \includegraphics[width=\linewidth]{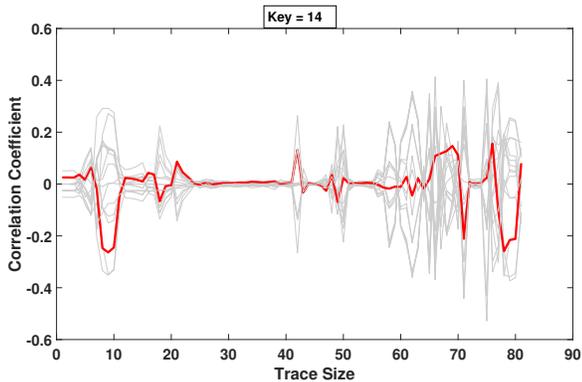}
\caption{Unsuccessful CPA Attack on hybrid adiabatic-MTJ based implementation of PRESENT S-box.}
\label{spfa_cpa}

\end{subfigure}

\caption{Correlation power analysis performed on both CMOS and hybrid adiabatic-MTJ implementation of PRESENT-80.}

\end{figure}

Previously we have shown our proposed hybrid adiabatic-MTJ based implementation of PRESENT to consume less energy when compared to CMOS. While reducing the energy consumption of a circuit we must also ensure the resilience of a circuit against side-channel attacks. The S-box of PRESENT will be the attack point in the Correlation Power Analysis (CPA) attack. The CPA attack is performed by following the steps described in \cite{wu2012measurement}. The simulation was performed at 12.5 MHz with a 10fF load. Practical CPA attacks usually require a large amount of traces to steal encryption keys. However, we are performing a simulation without electrical noise and therefore we require much fewer traces to steal the encryption key. In our attack, we have chosen 80 samples per clock period thus we will sample every 1ns. Using 5120 input traces, we were able to steal the encryption key in the CMOS-based design of PRESENT-80. Figure \ref{cmos_cpa} shows a successful CPA attack on a CMOS implementation of PRESENT-80.

While the CMOS key was revealed in 5120 traces, the hybrid adiabatic-MTJ implementation of the PRESENT S-box did not reveal the key in greater than 12,000 traces. Figure \ref{spfa_cpa} shows an unsuccessful CPA attack against the hybrid adiabatic-MTJ implemented PRESENT S-box. This case study demonstrates our circuits resistance against CPA attacks and shows it is a promising candidate to design secure and low-energy IoT devices. 

\section{Conclusion and Future Work}
\label{sec:conc}
A novel hybrid adiabatic-MTJ circuit was presented in this paper. The novel circuit provides substantial energy savings and is also resistant to Correlation Power Analysis Attacks. As a case study, we constructed one round of PRESENT and demonstrated that it consumed lower energy when compared to its CMOS counterpart. Furthermore, we have performed a Correlation Power Analysis Attack on both implementations of the S-box and determined that we could retrieve the key from the CMOS implementation but not from the hybrid adiabatic-MTJ implementation. As future work, a reliability analysis will need to be conducted to determine the usefulness of the circuit under a variety of variations. 

\label{conc}

\section*{Acknowledgment}
This work is partially supported by National Science Foundation CAREER Award No. 1845448.

\bibliographystyle{IEEEtran}
\bibliography{citations}

\begin{thebibliography}{10}
\providecommand{\url}[1]{#1}
\csname url@samestyle\endcsname
\providecommand{\newblock}{\relax}
\providecommand{\bibinfo}[2]{#2}
\providecommand{\BIBentrySTDinterwordspacing}{\spaceskip=0pt\relax}
\providecommand{\BIBentryALTinterwordstretchfactor}{4}
\providecommand{\BIBentryALTinterwordspacing}{\spaceskip=\fontdimen2\font plus
\BIBentryALTinterwordstretchfactor\fontdimen3\font minus
  \fontdimen4\font\relax}
\providecommand{\BIBforeignlanguage}[2]{{%
\expandafter\ifx\csname l@#1\endcsname\relax
\typeout{** WARNING: IEEEtran.bst: No hyphenation pattern has been}%
\typeout{** loaded for the language `#1'. Using the pattern for}%
\typeout{** the default language instead.}%
\else
\language=\csname l@#1\endcsname
\fi
#2}}
\providecommand{\BIBdecl}{\relax}
\BIBdecl

\bibitem{alioto2017internet}
M.~Alioto and M.~Shahghasemi, ``The internet of things on its edge: Trends
  toward its tipping point,'' \emph{IEEE Consumer Electronics Magazine},
  vol.~7, no.~1, pp. 77--87, 2017.

\bibitem{huai2008spin}
Y.~Huai \emph{et~al.}, ``Spin-transfer torque mram (stt-mram): Challenges and
  prospects,'' \emph{AAPPS bulletin}, vol.~18, no.~6, pp. 33--40, 2008.

\bibitem{deng2013low}
E.~Deng, Y.~Zhang, J.-O. Klein, D.~Ravelsona, C.~Chappert, and W.~Zhao, ``Low
  power magnetic full-adder based on spin transfer torque mram,'' \emph{IEEE
  transactions on magnetics}, vol.~49, no.~9, pp. 4982--4987, 2013.

\bibitem{kang2016low}
W.~Kang, W.~Lv, Y.~Zhang, and W.~Zhao, ``Low store power high-speed
  high-density nonvolatile sram design with spin hall effect-driven magnetic
  tunnel junctions,'' \emph{IEEE Transactions on Nanotechnology}, vol.~16,
  no.~1, pp. 148--154, 2016.

\bibitem{kang2015spintronics}
W.~Kang, Y.~Zhang, Z.~Wang, J.-O. Klein, C.~Chappert, D.~Ravelosona, G.~Wang,
  Y.~Zhang, and W.~Zhao, ``Spintronics: Emerging ultra-low-power circuits and
  systems beyond mos technology,'' \emph{ACM Journal on Emerging Technologies
  in Computing Systems (JETC)}, vol.~12, no.~2, pp. 1--42, 2015.

\bibitem{zhao2013synchronous}
W.~Zhao, M.~Moreau, E.~Deng, Y.~Zhang, J.-M. Portal, J.-O. Klein, M.~Bocquet,
  H.~Aziza, D.~Deleruyelle, C.~Muller \emph{et~al.}, ``Synchronous non-volatile
  logic gate design based on resistive switching memories,'' \emph{IEEE
  Transactions on Circuits and Systems I: Regular Papers}, vol.~61, no.~2, pp.
  443--454, 2013.

\bibitem{teichmann2011adiabatic}
P.~Teichmann, \emph{Adiabatic logic: future trend and system level
  perspective}.\hskip 1em plus 0.5em minus 0.4em\relax Springer Science \&
  Business Media, 2011, vol.~34.

\bibitem{kocher1999differential}
P.~Kocher, J.~Jaffe, and B.~Jun, ``Differential power analysis,'' in
  \emph{Annual international cryptology conference}.\hskip 1em plus 0.5em minus
  0.4em\relax Springer, 1999, pp. 388--397.

\bibitem{dhem1998practical}
J.-F. Dhem, F.~Koeune, P.-A. Leroux, P.~Mestr{\'e}, J.-J. Quisquater, and J.-L.
  Willems, ``A practical implementation of the timing attack,'' in
  \emph{International Conference on Smart Card Research and Advanced
  Applications}.\hskip 1em plus 0.5em minus 0.4em\relax Springer, 1998, pp.
  167--182.

\bibitem{wu2012measurement}
J.~Wu, Y.~Shi, and M.~Choi, ``Measurement and evaluation of power analysis
  attacks on asynchronous s-box,'' \emph{IEEE Transactions on Instrumentation
  and Measurement}, vol.~61, no.~10, pp. 2765--2775, 2012.

\bibitem{bogdanov2007present}
A.~Bogdanov, L.~R. Knudsen, G.~Leander, C.~Paar, A.~Poschmann, M.~J. Robshaw,
  Y.~Seurin, and C.~Vikkelsoe, ``Present: An ultra-lightweight block cipher,''
  in \emph{International workshop on cryptographic hardware and embedded
  systems}.\hskip 1em plus 0.5em minus 0.4em\relax Springer, 2007, pp.
  450--466.

\bibitem{moodera1995large}
J.~S. Moodera, L.~R. Kinder, T.~M. Wong, and R.~Meservey, ``Large
  magnetoresistance at room temperature in ferromagnetic thin film tunnel
  junctions,'' \emph{Physical review letters}, vol.~74, no.~16, p. 3273, 1995.

\bibitem{zand2016scalable}
R.~Zand, A.~Roohi, S.~Salehi, and R.~F. DeMara, ``Scalable adaptive spintronic
  reconfigurable logic using area-matched mtj design,'' \emph{IEEE Transactions
  on Circuits and Systems II: Express Briefs}, vol.~63, no.~7, pp. 678--682,
  2016.

\bibitem{behin2014computing}
B.~Behin-Aein, J.-P. Wang, and R.~Wiesendanger, ``Computing with spins and
  magnets,'' \emph{arXiv preprint arXiv:1411.6960}, 2014.

\bibitem{kent2010perpendicular}
A.~D. Kent, ``Perpendicular all the way,'' \emph{Nature materials}, vol.~9,
  no.~9, pp. 699--700, 2010.

\bibitem{gang2011high}
Y.~Gang, W.~Zhao, J.-O. Klein, C.~Chappert, and P.~Mazoyer, ``A
  high-reliability, low-power magnetic full adder,'' \emph{IEEE Transactions on
  Magnetics}, vol.~47, no.~11, pp. 4611--4616, 2011.

\bibitem{athas1994low}
W.~C. Athas, L.~J. Svensson, J.~G. Koller, N.~Tzartzanis, and E.~Y.-C. Chou,
  ``Low-power digital systems based on adiabatic-switching principles,''
  \emph{IEEE Transactions on Very Large Scale Integration (VLSI) Systems},
  vol.~2, no.~4, pp. 398--407, 1994.

\bibitem{kahleifeh20202}
Z.~Kahleifeh and H.~Thapliyal, ``2-phase energy-efficient secure positive
  feedback adiabatic logic for cpa-resistant iot devices,'' in \emph{2020 IEEE
  6th World Forum on Internet of Things (WF-IoT)}.\hskip 1em plus 0.5em minus
  0.4em\relax IEEE, 2020, pp. 1--5.

\bibitem{wang2016compact}
Y.~Wang, H.~Cai, L.~A. de~Barros~Naviner, Y.~Zhang, X.~Zhao, E.~Deng, J.-O.
  Klein, and W.~Zhao, ``Compact model of dielectric breakdown in spin-transfer
  torque magnetic tunnel junction,'' \emph{IEEE Transactions on Electron
  Devices}, vol.~63, no.~4, pp. 1762--1767, 2016.

\end{thebibliography}

\end{document}